\documentclass[a4paper,12pt,titlepage]{article}

\usepackage{jtb}
\usepackage{amsmath,amssymb}
\usepackage{bm}
\usepackage[dvips]{graphicx}
\usepackage{color}
\usepackage{comment}
\usepackage{multirow}

\title{Evolution of cooperation facilitated by reinforcement learning with adaptive aspiration levels}
\author{Shoma Tanabe$^{1}$ and Naoki Masuda$^{2,3,*}$\\
\\
$^1$ Faculty of Engineering, University of Tokyo \\
7-3-1 Hongo, Bunkyo, Tokyo 113-8656, Japan\\
$^2$ Graduate School of Information Science and Technology \\
University of Tokyo, 7-3-1 Hongo, Bunkyo, Tokyo 113-8656, Japan\\
$^3$ PRESTO, Japan Science and Technology Agency \\
4-1-8 Honcho, Kawaguchi, Saitama 332-0012, Japan\\
\\
$^*$ Corresponding author (masuda@mist.i.u-tokyo.ac.jp)\\
}
\date{\today}

\begin{document}
\bibliographystyle{jtb}
\maketitle
\begin{abstract}
Repeated interaction between individuals is the main mechanism for maintaining cooperation in social dilemma situations. 
Variants of tit-for-tat (repeating the previous action of the opponent) and the win-stay lose-shift strategy are known as strong competitors in iterated social dilemma games. 
On the other hand, real repeated interaction generally allows plasticity (i.e., learning) of individuals based on the experience of the past. 
Although plasticity is relevant to various biological phenomena, its role in repeated social dilemma games is relatively unexplored. 
In particular, if experience-based learning plays a key role in promotion and maintenance of cooperation, learners should evolve in the contest with nonlearners under selection pressure. 
By modeling players using a simple reinforcement learning model, we numerically show that learning enables the evolution of cooperation. 
We also show that numerically estimated adaptive dynamics appositely predict the outcome of evolutionary simulations. 
The analysis of the adaptive dynamics enables us to capture the obtained results as an affirmative example of the Baldwin effect, where learning accelerates the evolution to optimality. 
\end{abstract}
\section{Introduction}
The mechanisms of cooperation in social dilemma situations are a central topic in interdisciplinary research fields including evolutionary biology, ecology, economics, and sociology. 
As analyzed by the prisoner's dilemma (PD) game and its relatives, direct reciprocity is among the main known mechanisms underlying cooperative behavior \cite{Trivers1971,Axelrod1984book}. 
In direct reciprocity, iterated interaction between the same individuals motivates them to continue cooperating (C) rather than to defecting (D) to obtain momentarily large payoffs; defection would be negatively rewarded by the opponent player's retaliation in later rounds. 
Variants of the celebrated retaliatory strategy tit-for-tat (mimicking the opponent's action in the previous round) \cite{Nowak1992Nature_gtft} and a win-stay lose-shift strategy \cite{Kraines1989,Nowak1993Nature} are recognized as strong competitors in the iterated PD game. 

In the iterated games concerning direct reciprocity, it is natural to assume that players modify their strategies in response to their experiences in  past rounds. 
The tit-for-tat, its variants, and win-stay lose-shift strategies can be interpreted as examples of such learning strategies because the tit-for-tat, for example, implies that the player selects the action (i.e., C or D) depending on the result of the last round. 
A more sophisticated learning player of this kind exploits a longer history of the game for action selection (e.g., cooperate if the player and the opponent cooperated in the previous two rounds, and defect otherwise) \cite{Lindgren1991ALife}. 
Classes of other learning models include fictitious play and reinforcement learning \cite{Camerer2003book,Fudenberg1998book}. 
Learning apparently seems beneficial in iterated games because learning players are more flexible than nonlearning players. 

If learning is a key factor in promoting cooperation in real societies, the number of learning players should increase when a population evolves under selection pressure. 
However, the advantage of learners over nonlearners in evolutionary dynamics is elusive because a pair of learning players often results in mutual defection \cite{Macy1996SMR,Sandholm1996BioS,Posch1999RoyalB,Taiji1999PhysicaD,Macy2002pnas,MasudaOhtsuki2009BMB} and learning may be costly. 

%
%
The constructive roles of learning in the evolution of certain traits are collectively called the Baldwin effect (see Simpson, 1953; Turney \textit{et al.}, 1996; Weber \& D. J. Depew, 2003; Crispo, 2007; Badyaev, 2009 for reviews). 
Although earlier examples of the Baldwin effect are not necessarily founded on firm empirical evidence \cite{Simpson1953Evol,Weber2003book}, there exists a plethora of positive evidence of the Baldwin effect. 
Examples include fly's morphological developments \cite{Waddington1942Nature}, colonization of house finch in North America \cite{Badyaev2009PTRSB}, and persistence of coastal juncos \cite{Yeh2004AmNat}. 
%
%
In fact, the concept of the Baldwin effect differs by authors (see Simpson, 1953; Downes, 2003; Turney \textit{et al.}, 1996). 
Although earlier computational models suggest that learning accelerates evolution \cite{Hinton1987CS,Ancel1999JTB,Maynard-Smith1987Nature}, later theoretical and numerical studies suggest that learning either accelerates or decelerates evolution toward the optimum depending on the details of the models \cite{Ancel2000TPB,Dopazo2001BMB,Borenstein2006,Paenke2007AmNat,Paenke2009AI}. 
The advantage of learning in evolution is also nontrivial in this broader context. 

We numerically investigate the effect of learning on evolution in the iterated PD game. 
This question was explored in previous literature
%
%
(Suzuki \& Arita, 2004; also see Wang \textit{et al.}, 2008 for discussion).
%
%
Our emphasis in this study is to use a reinforcement learning model for the iterated PD game \cite{MasudaNakamura2011JTB} that is much simpler in terms of the number of plastic elements than the plastic look-up-table model adopted in \cite{Suzuki2004BioS}. 
In our model, players are satisfied with and persist in the current action when the obtained payoff is larger than a plastic threshold. 
Our model of players introduced in \cite{MasudaNakamura2011JTB} modifies those in \cite{Karandikar1998JET,Posch1999RoyalB,Macy2002pnas}. 
Via the stability analysis for nonlearning players, the numerical analysis of the discretized adaptive dynamics with nonlearning and learning players, and full evolutionary simulations, we show that learning is needed for a noncooperative population to evolve to be able to engage in mutual cooperation for wide parameter ranges. 
We also discuss our results in the context of the Baldwin effect. 
\section{Model}
\subsection{Iterated PD game}\label{sec:iterated PD}
We assume that each player plays the PD game against each of the other players in a population. 
In each round $t\ (t=1,\ 2,\ \ldots)$ within a generation, a player selects C or D without knowing the action (i.e., C or D) of the opponent player. 
The payoff to the focal player is defined by \vspace{-2mm}
\begin{eqnarray}
 \bordermatrix{
         &{\rm C}&{\rm D}\vspace{-2mm}\cr
  {\rm C}&     R &     S \vspace{-2mm}\cr
  {\rm D}&     T &     P 
 },
\label{eq:payoff matrix}
\end{eqnarray}
where $T>R>P>S$ and $R>(T+S)/2$. 
Equation~\eqref{eq:payoff matrix} represents the row player's payoff. 
The payoff to the opponent (column player) is defined likewise; the PD game is symmetric. 
Because $T>R$ and $P>S$, mutual defection is the only Nash equilibrium of the single-shot PD game. 

However, players may continue mutual cooperation for their own benefits in the iterated PD game \cite{Trivers1971,Axelrod1984book}. 
We denote the number of rounds per generation by $t_{\max}$. 
Technically, the Nash equilibrium of the iterated PD game is perpetual mutual defection if the players know $t_{\max}$ beforehand. 
The number of rounds is often randomized to avoid this effect \cite{Axelrod1984book}. 
To simplify the analysis, we assume that the players are unaware of the fixed value of $t_{\max}$. 

Earlier studies identified tit-for-tat, which involves imitating the previous action of the opponent, as a strong strategy in the iterated PD game when various strategies coexist in a population \cite{Trivers1971,Axelrod1984book}. 
However, later studies showed that tit-for-tat is not robust against error and that alternative strategies such as generous tit-for-tat \cite{Nowak1992Nature_gtft} and Pavlov \cite{Kraines1989,Nowak1993Nature} are strong competitors in the iterated PD game with error. 
By definition, a Pavlov player receiving payoff $T$ or $R$ is satisfied and does not change the action in the next round, whereas the same player receiving payoff $P$ or $S$ is dissatisfied and flip the action. 
A population composed of Pavlov players, for example, realizes mutual cooperation such that a player gains approximately $R$ per round. 
\subsection{Reinforcement learning}\label{sec:reinforcement learning}
Intuitively, the ability to learn may seem to be an advantageous trait in the iterated PD game if the cost of learning is negligible. 
However, this is generally not the case. 
A pair of learning players often ends up with mutual defection unless a learning algorithm is carefully designed \cite{Macy1996SMR,Sandholm1996BioS,Posch1999RoyalB,Taiji1999PhysicaD,Macy2002pnas,MasudaOhtsuki2009BMB}. 
Learning requires trial and error, i.e., the exploration of unknown behavioral patterns as well as the exploitation of known advantageous behavioral patterns. 
Exploratory behavior of a learning player may look just random to opponents, and it is rational to defect against random-looking players. 

To compromise the possibility of mutual cooperation, the simplicity of the learning algorithm, and the biological plausibility of the model as compared to some other learning algorithms, 
we use a variant of the Bush--Mosteller (BM) reinforcement learning model \cite{MasudaNakamura2011JTB}. 
This model modifies the models in the previous literature \cite{Karandikar1998JET,Posch1999RoyalB,Macy2002pnas} such that players learn to mutually cooperate for wide parameter ranges. 

In round $t$, the cooperability of the learning player is given by the probability $p_t$. 
We update $p_t$ using the results of the single-shot PD game as follows: 
\begin{eqnarray}
 p_{t+1}=\left \{
 \begin{array}{ll}
  p_t+(1-p_t)s_t & ({\rm action\ in\ round\ } t {\rm\ is\ C,\ and\ }s_t \geq 0),\vspace{-2mm}\cr
  p_t+p_ts_t     & ({\rm action\ in\ round\ } t {\rm\ is\ C,\ and\ }s_t < 0),   \vspace{-2mm}\cr
  p_t-p_ts_t     & ({\rm action\ in\ round\ } t {\rm\ is\ D,\ and\ }s_t \geq 0),\vspace{-2mm}\cr
  p_t-(1-p_t)s_t & ({\rm action\ in\ round\ } t {\rm\ is\ D,\ and\ }s_t < 0), 
 \end{array}
 \right .
\label{eq:p_t update}
\end{eqnarray}
where 
\begin{equation}
 s_t=\tanh [\beta(r_t-A_t)], 
\label{eq:s_t update}
\end{equation}
and $r_t \in \{R,T,S,P\}$ is the payoff to the player in round $t$. 
$s_t$ stands for the degree of satisfaction in round $t$. 
When $s_t$ is large, the player increases the probability of taking the current action in round $t+1$. 
For example, the third line in Eq.~\eqref{eq:p_t update} indicates that the player decreases the probability of cooperation $p_t$ because selecting D in round $t$ has yielded a satisfactory outcome. 
In addition, we assume that the player misimplements the action with a small probability $\epsilon$ such that the player in fact cooperates with probability $(1-2\epsilon)p_t+\epsilon$ in round $t$. 
Equations~\eqref{eq:p_t update} and \eqref{eq:s_t update} indicate that the player is satisfied with the current situation if the obtained payoff $r_t$ is larger than the so-called aspiration level $A_t$. 
Otherwise, the player is motivated to flip the action. 
$\beta$ controls the sensitivity in the plasticity of $p_t$. 
If $\beta=0$, $s_t=0$ for any $t$ such that $p_t$ is constant. 
If $\beta=\infty$, $s_t=1$ or $-1$ for any $t$ such that $p_{t+1}=1$ or 0. 

Unless otherwise stated, we set the initial condition to $p_1=0$, i.e., the player defects in round 1. 
This value of $p_1$ is the most adverse to mutual cooperation. We will confirm in
Secs.~\ref{sec:Nash h=0} and \ref{sec:evolutionary simulations} that our main results are qualitatively the same if we set $p_1=1$. 

The dynamics of the aspiration level are given by 
\begin{equation}
 A_{t+1}=(1-h)A_t+hr_t,
 \label{eq:A_t update}
\end{equation}
where $h$ represents the learning rate. 
If $h=0$, $A_t$ is constant, and the model is equivalent to the classical BM model. 
If $h=1$, the player compares the current payoff and the payoff obtained in the last round to determine $s_t$. 
In our previous work, we showed that mutual cooperation is established among the players only after $t_{\max}=100$ rounds if $\beta$ is large and $h$ is small \cite{MasudaNakamura2011JTB}. 
In the numerical simulations, we set $\beta=3$, which is large enough to support mutual cooperation if other conditions, such as small $h$ and small $\epsilon$, are met. 

We remark that the initial condition $A_1$ is a key parameter to characterize the player. 
\subsection{Evolutionary dynamics}\label{sec:evolutionary dynamics}
We set the number of players in the population to $N=500$. 
In a single generation, each player $i$ plays the iterated PD game with $t_{\max}=200$ against all the other players. 
We always reset $p_t$ and $A_t$ to $p_1$ and $A_1$ when a player starts the iterated PD game with a new opponent. 
The single generation payoff $\overline {r}_i$ $(\in [S,T])$ is equal to the summation of the payoff obtained by playing against $N-1$ players, which is divided by $(N-1)t_{\max}$. 

After the single generation payoffs to all the players are determined, we select two players $i$ and $j$ with equal probability for strategy update. 
We use the Fermi rule \cite{Szabo1998,Traulsen2006} in which player $i$ adopts $j$'s $A_1$ and $h$ values in the next generation with probability $1/\left[1+\exp \left(\tilde \beta({\overline r}_i-{\overline r}_j) \right)\right]$, and player $j$ adopts $i$'s parameter values, otherwise. 
We set $\tilde \beta=1$. 
To account for mutation, we assume that after strategy update, $A_1$ and $h$ of the adopter are displaced by random small values obeying the uniform density on $[-\Delta_{A_1},\Delta_{A_1}]$ and $[-\Delta_h,\Delta_h]$, respectively. 
If the displaced $h$ exceeds 1 or is negative, we reset $h$ to 1 or 0, respectively. 
However, the resetting  seldom occurs in our evolutionary simulations. 

The phenotype of a player in round $t$ is specified by $p_t$ and $A_t$. 
It should be noted that $p_t$ and $A_t$ are not inherited over generations. 
In other words, the natural selection operates on the capacity to learn (i.e., $h$) but not on the acquired behavior (i.e., $p_t$ and $A_t$). 
Because we let $\beta$ in Eq.~\eqref{eq:s_t update} to be relatively large to realize mutual cooperation \cite{MasudaNakamura2011JTB}, $p_t$ is sensitive to the excess payoff relative to $A_t$ in the sense that $p_t$ is close to $0$ or $1$ unless $r_t$ is close to $A_t$. 
Therefore, $p_t$ is similar to the probability of cooperation conditioned on the outcome of the PD game in the previous round. 
When we use the term learning in the following, we exclusively refer to that induced by $h$ in the iterated PD game. A positive value of $h$ directly raises the plasticity of $A_t$ and indirectly controls that of $p_t$. 
\section{Results}
\subsection{Nash equilibria when without learning}\label{sec:Nash h=0}
To show that learning is necessary for the emergence of cooperation, we start by analyzing the competition between players that do not learn. 
With learning rate $h$ equal to zero, the aspiration level is fixed over rounds (i.e., $A_t=A_1$, $t \geq 1$). 
For the sake of analysis, we set $\beta = \infty$. 
Then, Eqs.~\eqref{eq:p_t update} and \eqref{eq:s_t update} imply that the player persists in the current action (i.e., $(p_t,p_{t+1})= (0,0)$ or $(1,1)$) if $r_t-A_t \geq 0$ and flips the action (i.e., $(p_t,p_{t+1})=(0,1)$ or $(1,0)$) otherwise. 
When $A_t$ is fixed, there are five strategies: 
\begin{itemize}
\item Strategy st1 is defined by $A_t<S$. Except for the action misimplemantation, an st1 player always cooperates or always defects, depending on the action in the first round. 
\item Strategy st2 is defined by $S<A_t<P$. An st2 player does not flip the action unless $r_t=S$. 
\item Strategy st3 is defined by $P<A_t<R$. An st3 player does not flip the action if mutual cooperation or unilateral defection is realized. It is equivalent to Pavlov, which is a strong competitor in the iterated PD game \cite{Kraines1989,Nowak1993Nature}. 
\item Strategy st4 is defined by $R<A_t<T$. An st4 player flips the action unless $r_t=T$. 
\item Strategy st5 is defined by $T<A_t$. An st5 player flips the action in every round except when the player misimplements the action. 
\end{itemize}

In Table~\ref{table:tab1}, the average payoff to a nonlearning (i.e., $h=0$) player (row player) playing against another nonlearning player (column player) is shown for $0<\epsilon \ll 1$ and $t_{\max} = \infty$. 
For example, st1 playing against st2 obtains $(R+3P+2S)/6$ per round on an average. 
The results shown in Table~\ref{table:tab1} are a subset of those obtained in \cite{Nowak1995JMB} (see Appendix A for details). 
Table~\ref{table:tab1} indicates that st3 is a Nash equilibrium when the five strategies are considered. 
In particular, st3 playing against another st3 realizes mutual cooperation and obtains the largest average payoff per round $R$. 
Therefore, a unanimous population composed of st3 players represents a eusocial situation. 
Mutual cooperation is not realized by any other combination of two players. 

Table~\ref{table:tab1} indicates that st2 is also a Nash equilibrium when $3P>R+2S$. 
In addition, although st4 is not a Nash equilibrium, a homogeneous population composed of st4 players is resistant to invasion by st3 in evolutionary situations because st4 gains a larger payoff than st3 does when playing against an st4 opponent. 

\begin{figure}[htbp]
  \begin{center}
  \includegraphics[width=\linewidth]{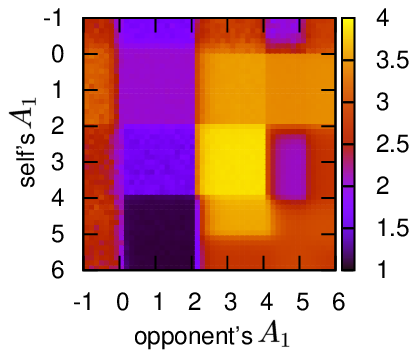}
  \end{center}
 \caption{
Average payoff to a nonlearning player (row player) playing against an opponent nonlearning player (column player). 
We set $\beta=3$, $p_1=0$, $\epsilon=0.02$, $h=0$, $t_{\max}=200$, $R=4,\ T=5,\ S=0,$ and $P=2$. 
}
 \label{fig:nl vs nl}
\end{figure}

To test the robustness of the results shown in Table~\ref{table:tab1}, we set $\beta=3,\ \epsilon=0.02$, and $h=0$, and numerically calculate the payoff averaged over $t_{\max}$ generations to different nonlearning players with different fixed aspiration levels $A_1$. 
We also set $R=4,\ T=5,\ S=0,$ and $P=2$ in this and the following numerical simulations. 
The average payoff to a nonlearner playing against another nonlearner is shown for $p_1=0$ and $t_{\max}=200$ in Fig.~\ref{fig:nl vs nl}. 
The presented values are averages over 100 trials for each pair of $A_1$ values. 
The results shown in Fig.~\ref{fig:nl vs nl} are qualitatively the same as those shown in Table~\ref{table:tab1}. 
We also confirmed that the results hardly change for $(p_1,t_{\max})=(0,2000),\ (1,200),$ and $(1,2000)$. 
\subsection{Possibility of mutual cooperation via learning}\label{sec:mutual C}
If $h>0,$ players different from st3 may adjust $A_t$ until $P<A_t<R$ is satisfied such that they learn to behave as Pavlov. 
Therefore, learning may play a constructive role in the evolution of mutual cooperation. 
In fact, this is not always the case; $\epsilon>0$ is a necessary condition for mutual cooperation to evolve. 

\begin{figure}[htbp]
 \begin{minipage}{0.495\linewidth}
  \begin{center}
  \includegraphics[width=\linewidth]{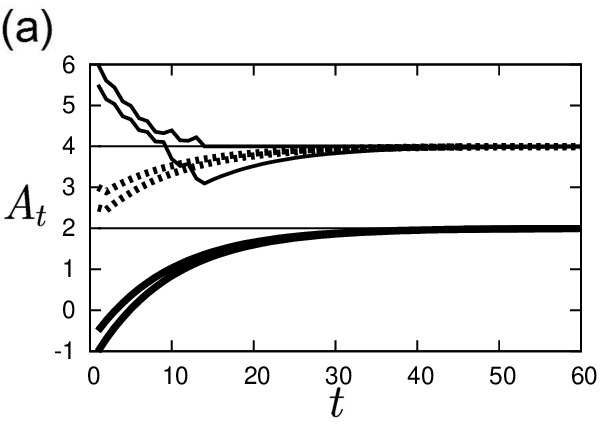}
  \end{center}
 \end{minipage}
 \begin{minipage}{0.495\linewidth}
  \begin{center}
  \includegraphics[width=\linewidth]{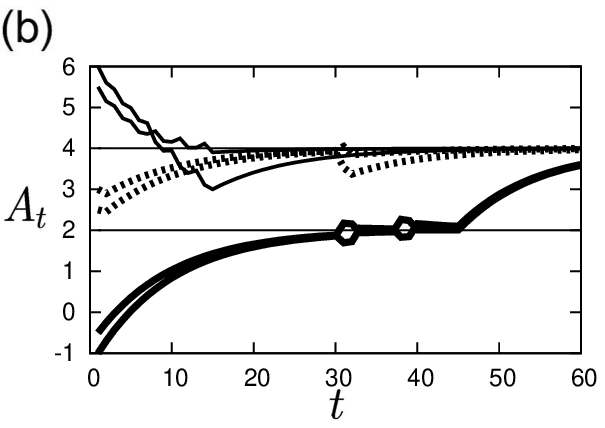}
  \end{center}
 \end{minipage}
 \caption{
Behavior of a pair of learning players. 
We set $\beta=3$, $p_1=0$, and $h=0.1$. 
(a) Example time courses of the aspiration level for a pair of players when $\epsilon=0$. 
The horizontal lines represent $A_t=P=2$ and $A_t=R=4$. 
We set $A_1$ for the two players to $-1$ and $-0.5$ (thick lines), $2.5$ and $3$ (dotted lines), and $5.5$ and $6$ (medium lines). 
(b) Example time courses of the aspiration level when $\epsilon=0.02$. 
We set $A_1$ as in (a). 
For each pair, one of the two players is assumed to misimplement the action
to cooperate in round $30$ (thick lines), to defect in round $30$ (dotted lines), or to cooperate in round $7$ (medium lines). 
}
 \label{fig:l vs l}
\end{figure}

To explain this point, we set $\beta=3$ and $h=0.1$, and numerically examine the behavior of a pair of players. 
Typical time courses of the aspiration level for a pair of learning players over rounds without action misimplementation (i.e., $\epsilon=0$) are shown in Fig.~\ref{fig:l vs l}(a). 
Each of the three pairs with close $A_1$ values represents a pair of st1 (thick lines), st3 (dotted lines), and st5 players (medium lines), respectively. 
We used different values of $A_1$ for each pair for the clarity of the figure; making $A_1$ equal for two players does not qualitatively change the results. 
The thick lines in Fig.~\ref{fig:l vs l}(a) indicate that the two st1 players playing with each other are satisfied with payoff $P=2$ obtained by mutual defection. 
Therefore, their aspiration levels converge to $A_t=P$. 
The results would be the same if we start from a pair of st2 players or a combination of an st1 player and an st2 player. 
A pair of st3 players begin mutual cooperation from the second round, and their $A_t$ values converge to $R=4$ (dotted lines). 
Mutual cooperation is also realized if the two players are initially either st4 or st5, although some rounds are required before the players mutually cooperate (medium lines). 

Although two learning players having $A_t<P$ do not end up with mutual cooperation when $\epsilon=0$, the action misimplementation (i.e., $\epsilon>0$) can trigger a shift from mutual defection to mutual cooperation. 
Artificially generated time courses in the presence of action implementation are shown in Fig.~\ref{fig:l vs l}(b) for expository purposes. 
Until the intended action is misimplemented ($1 \leq t \leq 29$ in Fig.~\ref{fig:l vs l}(b)), two players starting with $A_1<P$ keep mutual defection (thick lines). 
When $A_t$ has sufficiently approached $P$, we assume that one player misimplements the action ($t=30$). 
Then, the $A_t$ values of both players cross $P$ from below within a couple of rounds such that the players start to behave as Pavlov and mutually cooperate. 
The possibility of mutual cooperation through this mechanism is sensitive to the value of $h$. 
Two players starting with $A_1<P$ end up with $A_t>P$ owing to the action misimplementation when $(P-S)/(T-P)<(1-h)^2$ (see Appendix B for derivation). 
When $T=5,\ S=0$, and $P=2$, this condition yields $0<h<1-\sqrt{2/3} \approx 0.184$ for an arbitrary value of $R$. 
Then, the $A_t$ values of the two players converge to $R$. 

The $A_t$ values also converge to $R$ when we start with a pair of st3 players (dotted lines in Fig.~\ref{fig:l vs l}(b)) and a pair of st4 or st5 players (medium lines in Fig.~\ref{fig:l vs l}(b)). 
This is because mutual cooperation is stable against action misimplementation; if one player turns into D by action misimplementation in round $t$, both players defect in round $t+1$ and cooperate in round $t+2$, if the actions are not misimplemented in rounds $t+1$ and $t+2$. 
This event sequence is likely unless $\epsilon$ is large. 
\subsection{Adaptive dynamics}\label{sec:adaptive dynamics}
In the evolutionary numerical simulations that we will describe in Sec.~\ref{sec:evolutionary simulations}, we allow the initial aspiration level $A_1$ and learning rate $h$ to mutate (Sec.~\ref{sec:evolutionary dynamics}). 
If the distribution of $A_1$ and that of $h$ for an evolving population are single peaked and sufficiently localized, we can grasp the evolutionary dynamics for a population by tracking the dynamics of the population averages of $A_1$ and $h$, denoted by $\overline {A_1}$ and $\overline h$, respectively. 
In the extreme case in which all the players share identical values of $A_1$ and $h$, the instantaneous dynamics of $\overline {A_1}$ and $\overline h$ are captured by adaptive dynamics \cite{Metz96chapter,Hofbauer1998book,Hofbauer2003BAMS,Doebeli2004Science}. 
Adaptive dynamics reveal the possibility for mutants with a slightly deviated parameter value to invade a homogeneous resident population. 
In this section, we numerically examine two-dimensional adaptive dynamics with respect to $\overline {A_1}$ and $\overline h$ to foresee the evolutionary simulations carried out in Sec.~\ref{sec:evolutionary simulations}. 

In this and the following sections, we set $\beta=3,\ p_1=0,\ \epsilon=0.02$, and $t_{\max}=200$ unless otherwise stated. 
Consider a homogeneous population of players sharing the parameter values $\overline {A_1}=A_1$ and $\overline h=h$. 
A mutant player with aspiration level $A_1^{\prime}$ and learning rate $h^{\prime}$ can invade the population if 
\begin{equation}
 \pi[s^\prime,s]-\pi[s,s]>0
 \label{eq:pi ieq1}
\end{equation}
or
\begin{equation}
 \pi[s^\prime,s]-\pi[s,s]=0\ {\rm and}\ \pi[s^\prime,s^\prime]-\pi[s,s^\prime]>0,
 \label{eq:pi ieq2}
\end{equation}
where $s=(A_1,h)$ and $s^\prime =(A_1^\prime,h^\prime)$ are the strategies of the resident and mutant players, respectively, and $\pi[s_1,s_2]$ represents the average payoff of strategy $s_1$ when playing with strategy $s_2$. 
$s$ is ESS if the converse of Eq.~\eqref{eq:pi ieq1} or the converse of Eq.~\eqref{eq:pi ieq2} is satisfied. 
If Eq.~\eqref{eq:pi ieq1} or \eqref{eq:pi ieq2} is satisfied, the homogeneous population comprising strategy $s$ would evolve toward $s^\prime$. 
We numerically calculate $\pi[s^\prime,s]-\pi[s,s]$, where $s^\prime=(A_1+0.2,h),\ (A_1-0.2,h),\ (A_1,h+0.02),$ and $(A_1,h-0.02)$. 
We confine $s^\prime$ in the neighborhood of $s$ because the amount of mutation for $A_1$ and $h$ is assumed to be small. 
Examining $\pi[s^\prime,s]-\pi[s,s]$ corresponds to looking at the discretized adaptive dynamics, i.e., the discretized derivative of $\pi[s^\prime,s]$ with respect to $s^\prime$ at $s^\prime=s$. 

\begin{figure}[htbp]
  \begin{center}
   \includegraphics[width=\linewidth]{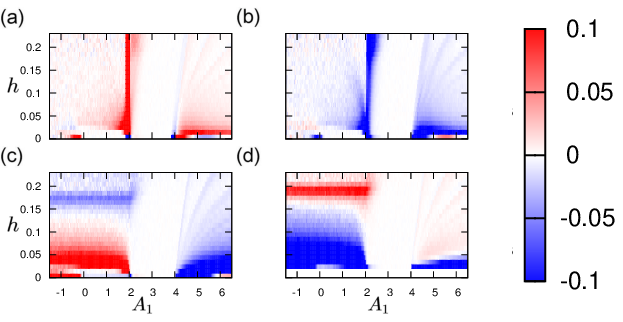}
  \end{center}
 \begin{minipage}{0.4\linewidth}
  \begin{center}
  \includegraphics[width=\linewidth]{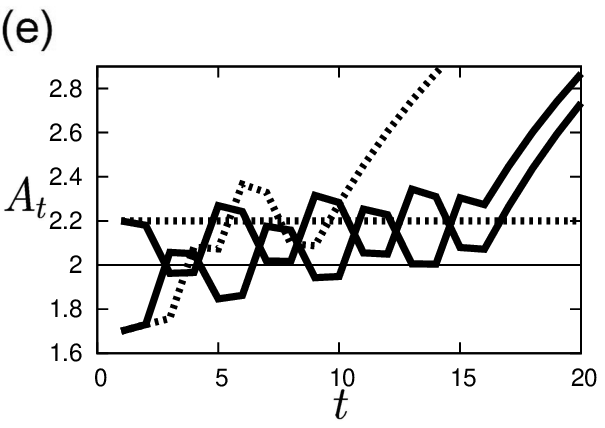}
  \end{center}
 \end{minipage}
 \caption{
Discretized adaptive dynamics when $\epsilon=0.02$. 
Plotted is $\pi[s^\prime,s]-\pi[s,s]$, where $s=(A_1,h)$ and (a) $s^\prime=(A_1+0.2,h)$, (b) $s^\prime=(A_1-0.2,h)$, (c) $s^\prime=(A_1,h+0.02)$, and (d) $s^\prime=(A_1,h-0.02)$. 
(e) Example time courses of a pair of players having $(A_1,h)=(1.7,0.1)$ and $(A_1,h)=(2.2,0.1)$ (solid lines), and $(A_1,h)=(1.7,0.1)$ and $(A_1,h)=(2.2,0)$ (dotted lines). 
The horizontal line indicates $A_t=P=2$.
We set $\beta=3$, $p_1=0$, and $t_{\max}=200$.
}
 \label{fig:adaptive dynamics}
\end{figure}

For various values of $A_1$ and $h$, $\pi[s^\prime,s]-\pi[s,s]$ is shown in Fig.~\ref{fig:adaptive dynamics}. 
The plotted values are averages over $10^4$ runs for any $s$. 
In Fig.~\ref{fig:adaptive dynamics}(a), $s^\prime=(A_1+0.2,h)$ obtains a larger payoff than $s=(A_1,h)$ in the red region. 
In this region, $s^\prime$ would invade a homogeneous resident population of $s$ such that $\overline{A_1}$ increases. 
In contrast, $s^\prime$ obtains a smaller payoff than $s$ does in the blue region. 
Figure~\ref{fig:adaptive dynamics}(a) indicates that, if learning is prohibited (i.e., $h=0$), the population starting from $A_1=0$, for example, is expected to evolve such that $\overline{A_1}$ increases, but only up to $A_1 \approx P=2$. 
Therefore, a population does not evolve from st2 to st3 without learning. 

Figure~\ref{fig:adaptive dynamics}(b), which reveals the possibility of invasion by mutant $s^\prime=(A_1-0.2,h)$ in the resident population of $s$, is a sign flipped version of Fig.~\ref{fig:adaptive dynamics}(a) in most parameter regions. 
Nevertheless, neither the mutants with $A_1^\prime=A_1+0.2$ nor the ones with $A_1^\prime=A_1-0.2$ invade the resident population (i.e., parameter regions colored in blue in both Figs.~\ref{fig:adaptive dynamics}(a) and \ref{fig:adaptive dynamics}(b)) for $(A_1,h) \approx (2,0)$ and along a bent line passing through $(A_1,h) \approx (4,0)$ and $(A_1,h) \approx (4.3,0.2)$. 
These regions constitute singular points of the adaptive dynamics and serve as repellers. 
In other words, $\overline{A_1}$ does not pass through $\approx P=2$ for $h=0$ and $\approx R=4$ for various values of $h$ in adaptive dynamics. 
The observations for $h=0$ that the homogeneous population of st2 is not invaded by st3 mutants, that of st3 is not invaded by st2 or st4 mutants, and that of st4 is not invaded by st3 mutants, are consistent with the results obtained in Sec.~\ref{sec:Nash h=0}. 

The possibility of invasion by mutant $s^\prime=(A_1,h+0.02)$ in the homogeneous population of $s=(A_1,h)$ is shown in Fig.~\ref{fig:adaptive dynamics}(c). 
The figure suggests that $\overline h$ would increase for a population of st1 players (i.e., $A_1<S=0$). 
Learning is preferred to nonlearning when $A_1<0$ for the following reason. 
As shown in Sec.~\ref{sec:mutual C}, when $h>0$ and $\epsilon>0$, $A_t$ increases until the players behave as Pavlov to mutually cooperate within a relatively small number of rounds (Fig.~\ref{fig:l vs l}(b)). 
In contrast, the players do not establish mutual cooperation when $h=0$ or $\epsilon=0$, as shown in Sec.~\ref{sec:Nash h=0} (Fig.~\ref{fig:l vs l}(a)). 
Figure~\ref{fig:adaptive dynamics}(c) indicates that $\overline h$ increases up to $\overline h \approx 0.15$. 
This value of $\overline h$ is consistent with the upper bound of $h$ for mutual cooperation to be possible, which was derived in Sec.~\ref{sec:mutual C}. 
Based on these results, $\overline h$ is expected to initially increase in evolutionary dynamics starting with a population of nonlearning st1 players. 
We refer to the stage of evolutionary dynamics in which $\overline h$ increases as stage 1. 
The existence of stage 1 is also supported by Fig.~\ref{fig:adaptive dynamics}(d) in which the mutant has $s^\prime=(A_1,h-0.02)$. 

After $\overline h$ has increased, Figs.~\ref{fig:adaptive dynamics}(a) and \ref{fig:adaptive dynamics}(b) imply that $\overline{A_1}$ increases to cross $P=2$. 
When $h>0$, a larger value of $A_1\ (<P)$ is beneficial because fewer rounds are required for such players to turn to Pavlov (i.e., $A_t>P$). 
Once $A_1$ exceeds $P$ for a majority of players, they earn a large average payoff $\approx R$ through mutual cooperation. 
We refer to the transition for learning players from a small $A_1$ corresponding to st1 or st2 to a large $A_1$ corresponding to st3 as stage 2. 
Figures~\ref{fig:adaptive dynamics}(a) and \ref{fig:adaptive dynamics}(b) indicate that the difference between $\pi[s^\prime,s]$ and $\pi[s,s]$ when $A_1$, $A_1+0.2<P$, $h>0$ is small, presumably because $s$ and $s^\prime$ are only slightly different in terms of the number of transient rounds before the entrance to $A_1>P$. 
Therefore, we expect that stage 2 occurs slowly in evolutionary dynamics. 

Although it is a minor phenomenon as compared to stages 1 and 2, a smaller $h$ is more beneficial on the boundary between st2 and st3 (i.e., $A_t \approx P=2$), as shown in Figs.~\ref{fig:adaptive dynamics}(c) and \ref{fig:adaptive dynamics}(d). 
For expository purposes, time courses of the iterated PD game between an st2 player and an st3 player are shown in Fig.~\ref{fig:adaptive dynamics}(e). 
As shown by the solid lines, the initial st3 player flips to st2 before establishing mutual cooperation if $h>0$. 
In fact, a nonlearning st3 player (i.e., $h=0$) realizes mutual cooperation with a learning st2 player in earlier rounds (dotted lines) than a learning st3 player does (solid lines). 
Therefore, in evolutionary dynamics, $\overline h$ in the vicinity of $A_t \approx P$ is expected to decrease. 
We refer to this transition as stage 3. 
It should be noted that stage 3 occurs in a narrow range of $A_1$ (i.e., $A_1<P$ and $A_1+0.2>P$ in Fig.~\ref{fig:adaptive dynamics}(a)). 

Through stages 1, 2, and 3, evolution from a defective population of nonlearning st1 players to a cooperative population of st3 players is logically possible. 
In contrast, the emergence of mutual cooperation is hampered if learning is prohibited. 

After stage 3, $\overline {A_1}$ would not evolve beyond $R$; $A_1 \approx R$ is a line of repellers in adaptive dynamics, as already explained in Figs.~\ref{fig:adaptive dynamics}(a) and \ref{fig:adaptive dynamics}(b). 
When $P<A_1<R$ (i.e., $2<A_1<4$), the mutant's payoff is indistinguishable from the resident's payoff unless $h$ is large (Figs.~\ref{fig:adaptive dynamics}(a)--(d)). 
Therefore, $\overline {A_1}$ and $\overline h$ would perform approximately unbiased diffusion. 
This implies that $\overline h$ that has decreased via stage 3 may increase again. 

When at least one of the two players is st4 or st5, a player with a larger $A_1$ is more advantageous than the opponent with a smaller $A_1$. 
This is because the former exploits the latter in early rounds. 
Nevertheless, these players do not obtain the average payoff as large as that for a pair of st3 players, which would start to mutually cooperate from the second round. 
Therefore, st3 is stable against invasion by st4 and vice versa. 

We predict that the learning rate would not eventually decrease to the small value in evolutionary simulations. 
In other words, the disadvantage of learning is too small to be evolutionarily relevant unless the cost of learning is explicitly incorporated. 

\begin{figure}[htbp]
  \begin{center}
   \includegraphics[width=\linewidth]{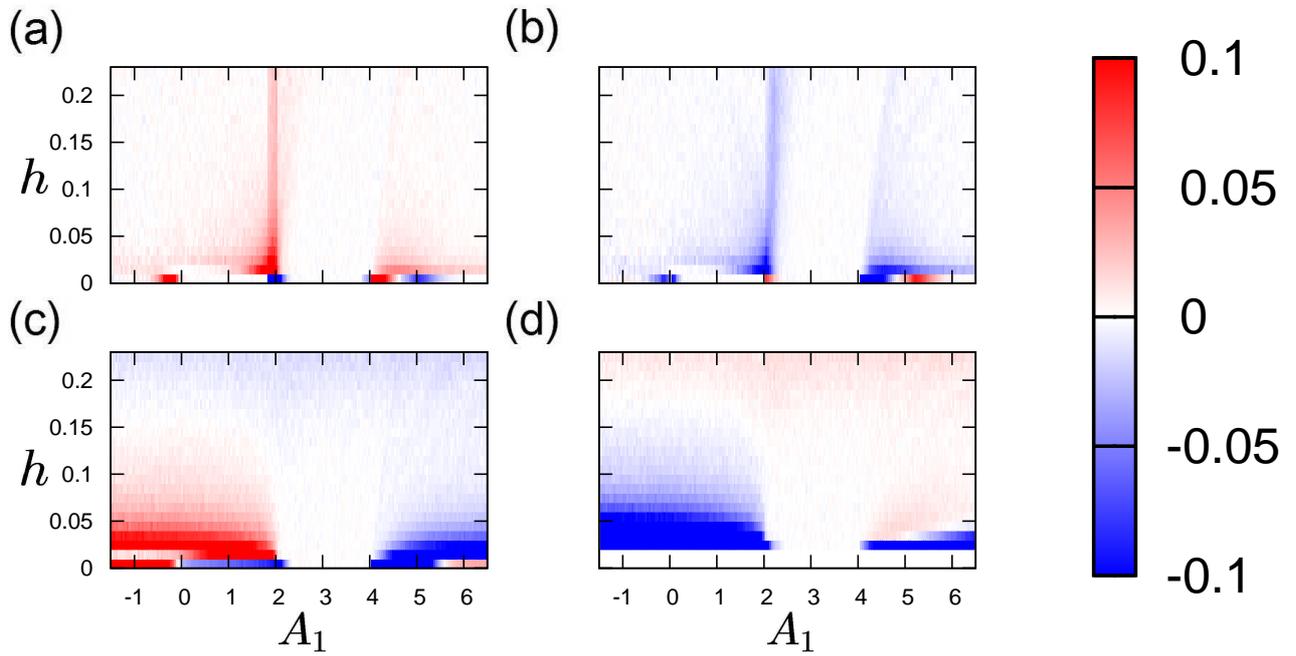}
  \end{center}
 \caption{
Discretized adaptive dynamics when $\epsilon=0.1$. 
Plotted is $\pi[s^\prime,s]-\pi[s,s]$, where $s=(A_1,h)$ and (a) $s^\prime=(A_1+0.2,h)$, (b) $s^\prime=(A_1-0.2,h)$, (c) $s^\prime=(A_1,h+0.02)$, and (d) $s^\prime=(A_1,h-0.02)$. 
}
 \label{fig:eps=0.1}
\end{figure}

\begin{figure}[htbp]
  \begin{center}
  \includegraphics[width=\linewidth]{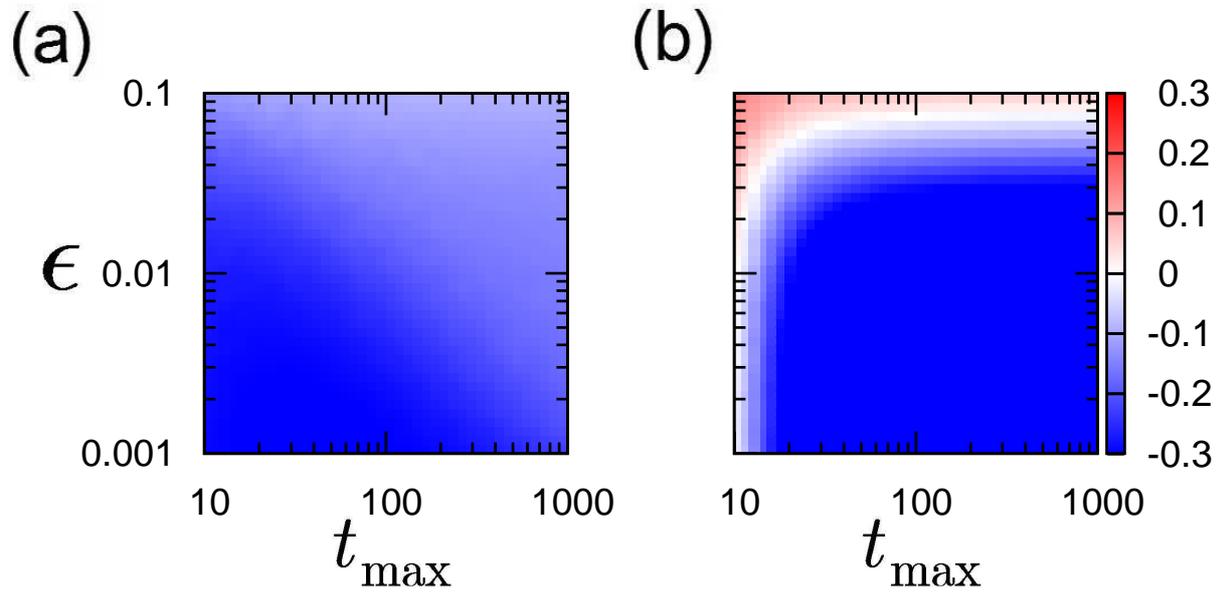}
  \end{center}
 \caption{
Effects of $t_{\max}$ and $\epsilon$ on the adaptive dynamics when $A_1 \approx P$ and $h=0$. 
We set $\beta=3$ and $p_1=0$.
(a) $\pi[(2.1,0),(1.9,0)]-\pi[(1.9,0),(1.9,0)]$. 
(b) $\pi[(1.9,0),(2.1,0)]-\pi[(2.1,0),(2.1,0)]$. 
}
 \label{fig:tmax eps}
\end{figure}

To assess the robustness of the results obtained from the adaptive dynamics, we reproduced Figs.~\ref{fig:adaptive dynamics}(a)--(d) with $\epsilon=0.05$ and $\epsilon=0.1$. 
The results for $\epsilon=0.05$ are qualitatively the same as those for $\epsilon=0.02$ (results not shown). 
The results for $\epsilon=0.1$ are different in some aspects from those for $\epsilon=0.02$ (Fig.~\ref{fig:eps=0.1}). 
Most notably, when $\epsilon=0.1$, st3 is no longer stable against invasion by st2 even without learning (i.e., $h=0$). 
Therefore, mutual cooperation would not be stable in evolutionary dynamics. 
In Figs.~\ref{fig:tmax eps}(a) and \ref{fig:tmax eps}(b), $\pi[s^\prime,s]-\pi[s,s]$ is shown for $(s,s^\prime)=((1.9,0),(2.1,0))$ and $(s,s^\prime)=((2.1,0),(1.9,0))$, respectively, for a variety of values of $t_{\max}$ and $\epsilon$. 
Figure~\ref{fig:tmax eps}(a) indicates that an st3 mutant does not invade the population of st2 residents for all the examined values of $t_{\max}$ and $\epsilon$. 
Figure~\ref{fig:tmax eps}(b) indicates that a population of st3 residents is resistant to invasion by st2 mutants when $t_{\max}$ is large and $\epsilon$ $(>0)$ is small. 
Nevertheless, st3 is stable for various values of $t_{\max}$ and $\epsilon$. 
Because stage 2 is hampered when $\epsilon=0$, $\epsilon$ must take an intermediate value for the learning-mediated mutual cooperation to emerge. 
\subsection{Evolutionary simulations}\label{sec:evolutionary simulations}
The results in Sec.~\ref{sec:adaptive dynamics} predict the presence of a learning mediated evolutionary route from a noncooperative population composed of st1 players to a cooperative population composed of st3 players. 
In this section, we carry out direct numerical simulations of the evolutionary dynamics using a population composed of $N=500$ players. 
We initially set $h=0$ and select $A_1$ for each player independently from the uniform density on $[S-1,S]$. 
Therefore, all the players are initially nonlearning st1. 
Refer to Sec.~\ref{sec:evolutionary dynamics} for details of the numerical setup. 

\begin{figure}[htbp]
 \begin{minipage}{0.56\linewidth}
  \begin{center}
  \includegraphics[width=\linewidth]{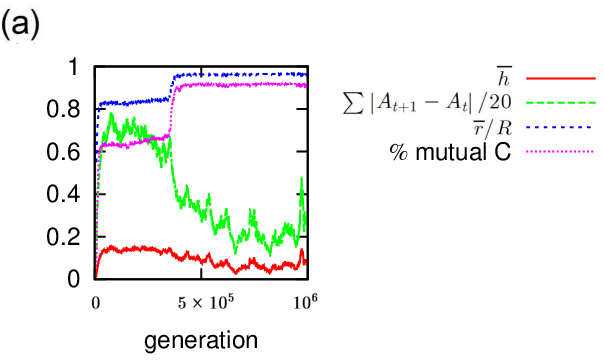}
  \end{center}
 \end{minipage}
 \begin{minipage}{0.43\linewidth}
  \begin{center}
  \includegraphics[width=\linewidth]{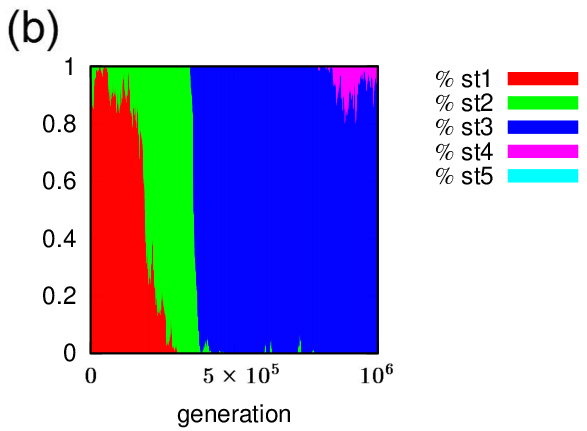}
  \end{center}
 \end{minipage}
 \begin{minipage}{0.495\linewidth}
  \begin{center}
  \includegraphics[width=\linewidth]{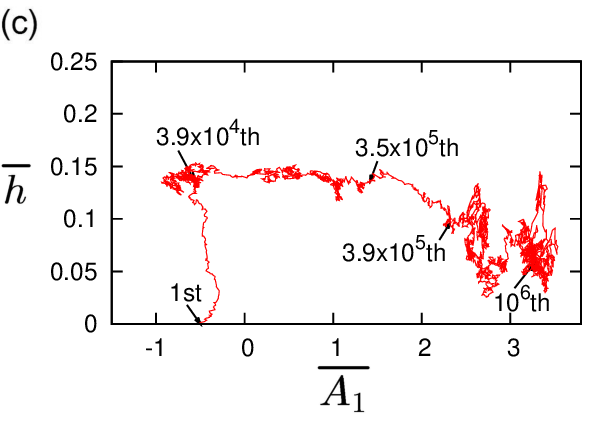}
  \end{center}
 \end{minipage}
 \caption{
Evolutionary dynamics in a population composed of learning players. 
We set $\beta=3$, $p_1=0$, $\epsilon=0.02$, $t_{\max}=200$, $\Delta_{A_1}=0.05$, and $\Delta_h=0.01$. 
(a) Time course of ${\overline h},\ {\sum_{t=1}^{t_{\max}-1}\left|A_{t+1}-A_t \right|}/20,\ {\overline r}/R$, and the fraction of mutual cooperation. 
(b) Time course of the fraction of initially st1, st2, st3, st4, and st5 players in the run shown in (a). 
(c) Sample trajectory of the population averages $\overline {A_1}$ and $\overline h$ in the run shown in (a). 
}
 \label{fig:evolutionary dynamics}
\end{figure}

The evolution of $\overline h$, the total amount of plasticity experienced in a generation, defined by $\sum_{t=1}^{t_{\max}-1} \left| A_{t+1}-A_t \right|$, $\overline r$, and the fraction of mutual cooperation for an example run with $\Delta_{A_1}=0.05$ and $\Delta_h=0.01$ are shown in Fig.~\ref{fig:evolutionary dynamics}(a). 
The average learning rate $\overline h$ and the total amount of plasticity rapidly increase until the $\approx 3.9\times10^4$th round. 
The payoff and the fraction of mutual cooperation also increase during this period because st1 players learn to behave as Pavlov when $h>0$. 
This period corresponds to stage 1 described in Sec.~\ref{sec:adaptive dynamics}. 
Then, the fraction of mutual cooperation and the total amount of plasticity gradually increase until the $\approx 3.5\times10^5$th round, corresponding to stage 2. 
In the $\approx 3.5\times10^5$th round, an st3 mutant emerges in the population mostly composed of st2 players and gains a larger payoff than st2 residents do. 
Then, st3 players rapidly replace st2 players in the population such that $\overline r/R$ and the fraction of mutual cooperation suddenly increase (Fig.~\ref{fig:evolutionary dynamics}(a)). 
This is because stable mutual cooperation between st3 players emerges in an early round, whereas that between st2 players emerges after $\approx 2/\epsilon$ rounds. 
The learning rate decreases almost at the same time, corresponding to stage 3. 
The time courses of the fractions of st1, st2, st3, st4, and st5 players corresponding to the run shown in Fig.~\ref{fig:evolutionary dynamics}(a) are shown in Fig.~\ref{fig:evolutionary dynamics}(b). 
For example, the fraction of the st1 player is defined by the fraction of players having $A_1<S$ and any value of $h$. 
Figure~\ref{fig:evolutionary dynamics}(b) indicates that the population initially composed of st1 players evolves to that of st3 players. 
The trajectory of $\overline {A_1}$ and $\overline h$ corresponding to the same run is shown in Fig.~\ref{fig:evolutionary dynamics}(c). 
Figure~\ref{fig:evolutionary dynamics}(c) is consistent with the scenario of the evolution of cooperation described in Sec.~\ref{sec:adaptive dynamics}. 
The population evolves from no cooperation to mutual cooperation via the three stages involving learning. 
After stage 3, $\overline {A_1}$ and $\overline h$ diffuse without a recognizable bias, which is also consistent with the results obtained in Sec.~\ref{sec:adaptive dynamics} (white regions in Figs.~\ref{fig:adaptive dynamics}(a)--(d)). 
However, it should be noted that the total amount of plasticity remains small after stage 3. 

To examine the robustness of the results, we carry out five runs of numerical simulations for each of the different parameter sets; we could not carry out more extensive numerical simulations because of the computational cost. 
We measure two quantities in each run. 
The first quantity is the number of generations necessary for $\overline h$ to exceed $0.1$ for the first time. 
We call this number the end of stage 1. 
The second quantity is the number of generations necessary for $\overline {A_1}$ to exceed $P$ for the first time. 
We call this number the end of stage 2. 
The ends of stages 1 and 2 with $\epsilon=0.02$, $\Delta_{A_1}=0.05$, and $\Delta_h=0.01$ are equal to $257\pm28\ (\times10^2)$ and $3763\pm1514\ (\times10^2)$, respectively, where the mean $\pm$ standard deviation on the basis of the five runs are indicated. 
Those with $\epsilon=0.05$, $\Delta_{A_1}=0.05$, and $\Delta_h=0.01$ are equal to $299\pm55\ (\times10^2)$ and $6951\pm4231\ (\times10^2)$. 
Those with $\epsilon=0.01$, $\Delta_{A_1}=0.05$, and $\Delta_h=0.01$ are equal to $265\pm31\ (\times10^2)$ and $3780\pm1547\ (\times10^2)$. 
Those with $\epsilon=0.02$, $\Delta_{A_1}=0.02$, and $\Delta_h=0.01$ are equal to $239\pm18\ (\times10^2)$ and $9151\pm6836\ (\times10^2)$. 
For this parameter set, one out of the five runs did not reach the end of stage 2 within $2\times10^6$ generations, such that the statistics are based on the other four runs.
Those with $\epsilon=0.02$, $\Delta_{A_1}=0.05$, and $\Delta_h=0.005$ are equal to $613\pm119\ (\times10^2)$ and $3504\pm999\ (\times10^2)$. 
Mutual cooperation evolves via learning (i.e., finite value of the end of stage 2 up to our numerical efforts) in most cases. 
When $\epsilon=0.05$, evolution to mutual cooperation is slower than when $\epsilon=0.02$. 
This may be because learning players having different values of $A_1$ turn into Pavlov (i.e., $A_t>P$) within a small number of rounds when $\epsilon$ is relatively large. 
Then, the payoff to different learning players would differ relatively little to weaken the selection pressure. 

\begin{figure}[htbp]
 \begin{center}
  \begin{minipage}{0.495\linewidth}
   \includegraphics[width=\linewidth]{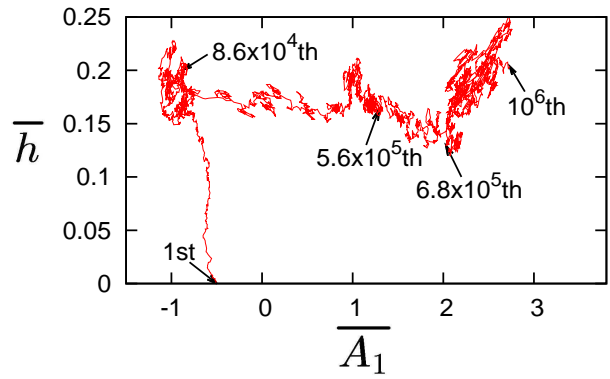}
  \end{minipage}
 \end{center}
 \caption{Sample trajectory of $\overline {A_1}$ and $\overline h$ when 
$\beta=3$, $p_1=1$, $\epsilon=0.02$, $t_{\max}=200$, $\Delta_{A_1}=0.05$, and $\Delta_h=0.01$. 
}
 \label{fig:p_1=1}
\end{figure}

We perform another robustness test. 
For the original parameter values $\epsilon=0.02$, $\Delta_{A_1}=0.05$, and $\Delta_h=0.01$, the trajectory of $\overline {A_1}$ and $\overline h$ obtained from a single run with $p_1=1$ is shown in Fig.~\ref{fig:p_1=1}. 
The results are qualitatively the same as those for $p_1=0$ (Fig.~\ref{fig:evolutionary dynamics}(c)) although establishment of cooperation takes a considerably larger number of generations when $p_1=1$ than when $p_1=0$. 
\subsection{Baldwin effect}\label{sec:Baldwin effect}
\begin{figure}[htbp]
 \begin{minipage}{0.495\linewidth}
  \begin{center}
  \includegraphics[width=\linewidth]{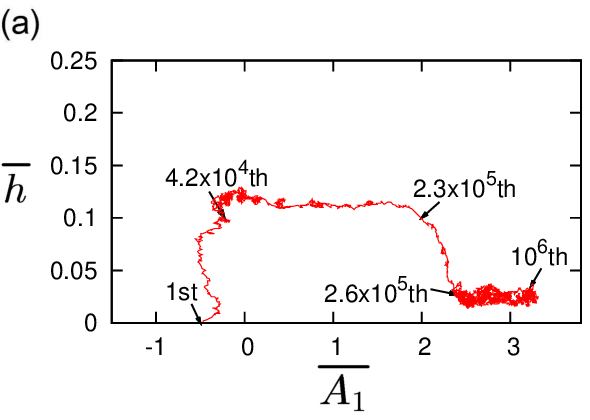}
  \end{center}
 \end{minipage}
 \begin{minipage}{0.495\linewidth}
  \begin{center}
  \includegraphics[width=\linewidth]{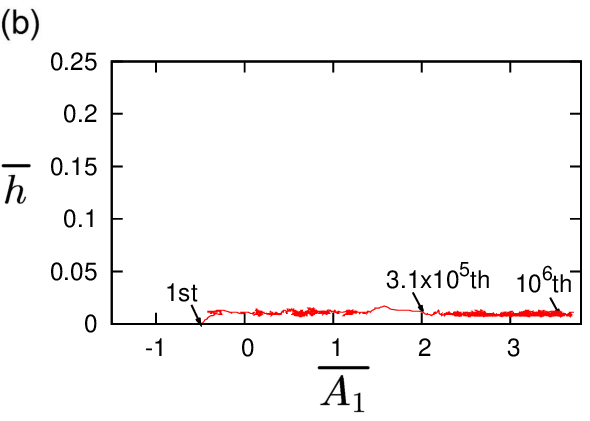}
  \end{center}
 \end{minipage}
 \begin{minipage}{0.495\linewidth}
  \begin{center}
  \includegraphics[width=\linewidth]{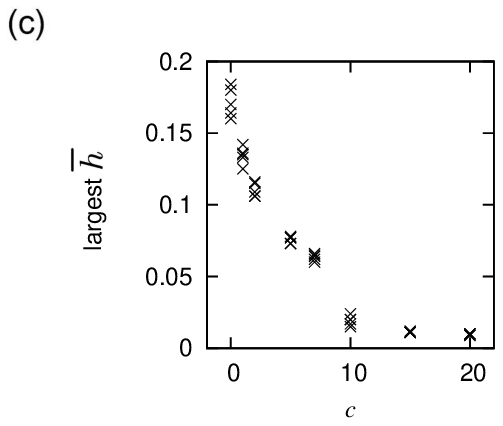}
  \end{center}
 \end{minipage}
 \begin{minipage}{0.495\linewidth}
  \begin{center}
  \includegraphics[width=\linewidth]{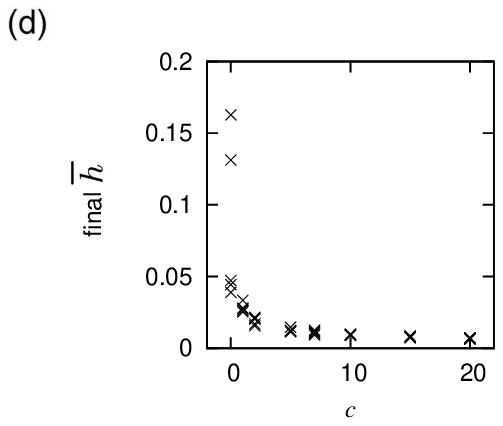}
  \end{center}
 \end{minipage}
 \begin{minipage}{0.495\linewidth}
  \begin{center}
  \includegraphics[width=\linewidth]{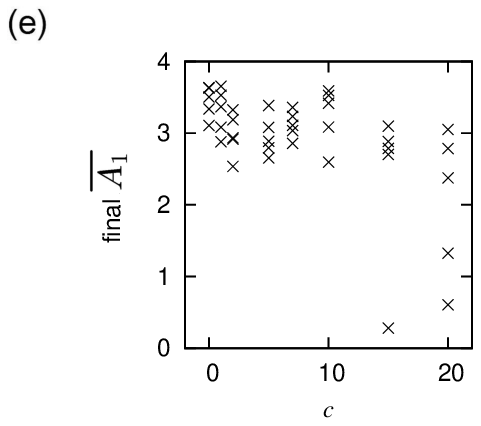}
  \end{center}
 \end{minipage}
 \caption{
(a, b) Sample trajectory of $\overline {A_1}$ and $\overline h$ under a linear cost of learning. We set (a) $c=1$ and (b) $c=10$. 
(c) Largest $\overline h$ in each of the five runs of $10^6$ generations for various values of $c$. A cross corresponds to a single run. 
(d) Average of $\overline h$ over the last $10^4$ generations in each run. 
(e) Average of $\overline {A_1}$ over the last $10^4$ generations in each run. 
}
 \label{fig:cost}
\end{figure}

If we assume an explicit cost of learning, the learning rate decreases after mutual cooperation is reached. 
%
%
An example time course of $\overline {A_1}$ and $\overline h$ when a linear cost $-ch$ is added to the single generation payoff to each player (Suzuki \& Arita 2004; see Ancel 1999, 2000 for a different implementation of the explicit learning cost), where $c=1$, is shown in Fig.~\ref{fig:cost}(a). 
The final value of $\overline h$ is smaller than that in the case without the learning cost (Fig.~\ref{fig:evolutionary dynamics}(c)).
The result shown in Fig.~\ref{fig:evolutionary dynamics}(d) is an example of the standard Baldwin effect in which the learning rate initially increases and then decreases \cite{Ancel2000TPB,Dopazo2001BMB,Borenstein2006,Paenke2007AmNat,Paenke2009AI}. 

We examine the robustness of the
observed Baldwin effect against the variation of $c$.
An example time course of $\overline {A_1}$ and $\overline h$ when $c=10$
is shown in Fig.~\ref{fig:cost}(b). 
As compared to when $c=1$ (Fig.~\ref{fig:cost}(a)),
$\overline h$ is smaller
throughout the evolution, and $\overline {A_1}$ increases more slowly.
Nevertheless, the population mostly
consists of st3 players in the end. We carried out
five runs for various values of $c$.
The largest $\overline h$ value in $10^6$ generations
is shown in Fig.~\ref{fig:cost}(c) for each run. 
The largest $\overline h$ value
decreases with $c$ because learning is costly for a large value of $c$.
The final value of $\overline h$, calculated as the average over the
last $10^4$ generations, is shown in Fig.~\ref{fig:cost}(d).
The final value of $\overline h$ is
considerably smaller than the largest value
(Fig.~\ref{fig:cost}(c)) for each $c$, indicative of stage 3 of the
Baldwin effect.
The final value of $\overline {A_1}$, calculated as the average over
the last $10^4$ generations, is plotted against $c$
in Fig.~\ref{fig:cost}(e). If this value is larger than $P=2$ and
smaller than $R=4$, we expect that the final population is
mostly composed of st3 players and that the Baldwin effect is operative.
Figure~\ref{fig:cost}(e) suggests that the Baldwin effect occurs in the five runs when $c<15$. When $c\ge 15$,
stage 1, i.e., the initial increase in $\overline h$, is often too
small
in magnitude such that stage 2 does not sometimes occur.
We conclude that the Baldwin effect occurs for a wide range of $c$.
\section{Discussion}
We have shown that reinforcement learning promotes the evolution of mutual cooperation in a population of players involved in the iterated PD game. 
Cooperation evolves under some conditions such as $3P>R+2S$, positive but not too large values of $\epsilon$, and $t_{\max}$ that is not too small. 
The present study is motivated by previous investigations of the Baldwin effect. 
Our results provide an example of the Baldwin effect in the form of a computational model of social behavior. 

To understand the behavior of our model analytically, writing
  down the Fokker--Planck equation for the joint density of $A_1$ and
  $h$ may be useful. Starting from the singular density at a small
  value of $A_1$ and $h=0$, we may be able to solve the Fokker--Planck
  equation numerically to track the evolution of the joint
  density to find the Baldwin
  effect. Alternatively, discretizing $A_1$ and $h$ and then
formulating a Markov chain on the discretized states may also be useful.
Nevertheless, we refrained from such analyses because we consider that they
eventually necessitates some numerical
  simulations and would not sufficiently advance the understanding of
  our numerical results.

The concept of the Baldwin effect is diverse \cite{Simpson1953Evol,Downes2003chap2,Turney1996EvolComput}. 
However, arguably, the most accepted variant of the Baldwin effect is formulated as a two-stage mechanism \cite{Simpson1953Evol,Godfrey2003chap3,Crispo2007Evol,Turney1996EvolComput}. 
In stage 1, plasticity increases because plastic individuals are better at finding the optimal behavior than nonplastic individuals. 
In stage 2, mutation makes the optimal behavior innate and decreases the plasticity of individuals. 
Mutants that play optimally from the outset of their life without plasticity and resident individuals that acquire the optimal behavior through plasticity are eventually equally efficient. 
Nevertheless, because of the cost of learning, the mutants overwhelm the residents via natural selection. 
Stage 2 is often called genetic assimilation. 

Stage 1 in our model corresponds to stage 1 of the standard Baldwin effect outlined above. 
In stage 2 in our model, $A_1$ increases such that the optimal behavior (i.e., mutual cooperation by turning into st3) becomes innate. 
Nevertheless, after stage 3 in our model in which the learning rate rapidly decreases, the learning rate starts to perform a random walk because the learning cost is marginal in our model (Fig.~\ref{fig:evolutionary dynamics}(c)). 
Therefore, the behavior of our model in stages 2, 3, and onward does not qualify as stage 2 of the standard Baldwin effect in which the learning rate decreases. 
With a modified model with an explicit learning cost, we showed that the learning rate decreases after stage 3 (Fig.~\ref{fig:evolutionary dynamics}(d)). 
In this case, our model naturally fits the framework of the Baldwin effect. 

In a previous computational model of the Baldwin effect, learning rates remain large when the optimal behavior dynamically changes owing to environmental fluctuations \cite{Ancel1999JTB}. 
In our model without an explicit cost of learning, the learning rate remains large for a different reason. 
In our model, the optimal parameter set (i.e., $A_1$ and $h$) does not fluctuate after sufficient generations. 
Instead, approximate optimality is realized for various parameter sets, i.e., any $P<A_1<R$ and $h \geq 0$. 
Therefore, the learning rate performs a random walk to occasionally visit large values (Fig.~\ref{fig:evolutionary dynamics}(c)). 

Godfrey--Smith points out three alternative reasons why stage 1 cannot be skipped in the two-stage mechanism of the Baldwin effect \cite{Godfrey2003chap3}. 
First, learning may provide a breathing space by which a population can survive long enough to transit to stage 2. 
This reason is irrelevant to our model because our model is not concerned with the survival of the population. 
The population size is fixed in our model such that the population always survives. 
Second, the preferred state may be accessible for learners but not for nonlearners. 
Although not explicitly stated in Godfrey--Smith (2003), this mechanism seems to be relevant to cases in which the fitness landscape does not depend on the configuration of the population. 
In our case, however, the fitness landscape depends on the fractions of the different types of players because the payoff to a player is affected by the strategies of the other players. 
Third, evolution may change the ``social ecology'' of the population such that learners are more advantageous than nonlearners, a phenomenon called niche construction in a broad sense. 
The social ecology implies a fitness landscape that depends on the configuration of the population. 
In our model, the social ecology evolves via learning of players. 
This third mechanism seems to be relevant to our model. 
Suppose a hypothetical population comprising st1 nonlearners except two st1 learners. 
For a focal st1 learner, the social ecology is such that there is one st1 learner and $N-2$ st1 nonlearners. 
If $\epsilon>0$, the focal st1 learner is likely to gain a payoff that is larger than an st1 nonlearner because the focal player learns to mutually cooperate with the other st1 learner, whereas an st1 nonlearner does not. 
The focal st1 learner would not overwhelm st1 nonlearners if the other st1 learner is absent in the social ecology. 

The main purpose of this study is to provide an evolutionary model of concrete social behavior in which learning plays a constructive role. 
We are not the first to achieve this end. 
Suzuki and Arita observed the Baldwin effect in the iterated PD game using different learning models \cite{Suzuki2004BioS}. 
In their model, the learning rate is assumed to be binary, and the player's strategy is specified by a look-up table that associates the action to take (i.e., C or D) with the actions of the previous two rounds of the two players. 
The entries of the look-up table dynamically change when the plasticity is in operation. 
They also considered the effects of meta--learning in which the player adapts how to update each entry of the look-up table. 
The main contribution of the present work relative to theirs is to provide a much simpler model in terms of the number of plastic parameters. 
In contrast, the learning rates and the range of parameters are continuous in our model, whereas they are mostly binary in their model. 
Our model may be amenable to real animals and facilitates a mechanistic understanding of evolutionary dynamics by the numerically calculated adaptive dynamics. 
Apart from the fixed parameters common to all the individuals, our players only have two parameters that are plastic within a generation, $p_t$ and $A_t$, and two parameters inherited across generations, $A_1$ and $h$. 
The results obtained from the adaptive dynamics predict those of direct evolutionary numerical simulations and provide an intuitive reason why learning promotes the emergence of mutual cooperation. 
In particular, we showed the necessity of the evolution of learning ability for cooperation by explicitly comparing the cases with and without learning. 
The combination of adaptive dynamics and evolutionary simulations may also be useful for analyzing the Baldwin effect in different models. 
\section*{Acknowledgements}
We thank Reiji Suzuki and Kohei Tamura for the valuable discussions. 
N.M. acknowledges the support provided through Grants-in-Aid for Scientific Research (Nos. 20760258 and 23681033, and Innovative Areas ``Systems Molecular Ethology''(No. 20115009)) from MEXT, Japan. 
\section*{Appendix A: Payoff to nonlearning players}
Nowak \textit{et al.} (1995) analyzed iterated matrix games between a pair of players that select an action (i.e., C or D) in response to the actions of the two players in the previous round. 
There are four combinations of the actions of the two players in the previous round, i.e., (C, C), (C, D), (D, C), and (D, D). 
Because a player assigns C or D to each of these possible outcomes in the previous round, there are 16 strategies $S_i$ $(0 \leq i \leq 15)$. 
In fact, st1, st2, st3, st4, and st5 in the present study are equivalent to $S_{12},\ S_8,\ S_9,\ S_1,$ and $S_3$ in \cite{Nowak1995JMB}, respectively. 

By calculating the steady state of the Markov chain with four states $R$, $T$, $S$, and $P$, Nowak \textit{et al.} (1995) calculated the average payoff to focal player $S_i$ playing against the opponent $S_j$ $(0 \leq i,\ j \leq 15)$ under a small probability of error in action implementation. 
Their assumption for the action misimplementation is slightly different from ours. 
We assumed that $\epsilon$ is the probability that each player independently misimplements the action, whereas only one of the two players may misimplement the action in a round in their model. 
Nevertheless, our model is equivalent to theirs in the limit $\epsilon \to 0$ if we set $\epsilon^{\prime}=2 \epsilon (1-\epsilon)$, where $\epsilon^{\prime}$ is the probability of action misimplementation in the sense of Nowak \textit{et al.} (1995). 
Therefore, our results shown in Table~\ref{table:tab1} are a corollary of their results. 
\section*{Appendix B: Upper bound of $h$ for st2 players to turn into Pavlov}
Given $\beta=\infty,\ 0<\epsilon \ll 1,\ h>0$, and $A_1<P$, $A_t$ of the two players, denoted by X and Y, are sufficiently close to $P$ when one player, which we assume to be Y without loss of generality, misimplements the action to select C for the first time in round $t \propto 2/\epsilon$. 
Without any further action misimplementation, X keeps D and Y flips to D in round $t+1$ because $A_{t+1}^{({\rm Y})}<P<A_{t+1}^{({\rm X})}$. 
In round $t+2$, X flips to C and Y keeps D. 
Therefore, we obtain $A_{t+3}^{({\rm X})}=hS+(1-h)A_{t+2}^{({\rm X})}$, $A_{t+2}^{({\rm X})}=hP+(1-h)A_{t+1}^{({\rm X})},$ $A_{t+1}^{({\rm X})}=hT+(1-h)A_t^{({\rm X})},$ and $A_t^{({\rm X})} \approx P$. 
Combining the four equations, we obtain 
\begin{equation}
 A_{t+3}^{({\rm X})}=(T-P)h^3-2(T-P)h^2+(T+S-2P)h+P. 
\end{equation}
Using $T>P>S$ and $0<h \leq 1$, we obtain the condition for X to become Pavlov in round $t+3$ as $A_{t+3}^{({\rm X})} > P$, i.e., 
\begin{equation}
 (1-h)^2 > \frac{P-S}{T-P}. 
 \label{eq:h iq1}
\end{equation}
The condition for Y to become Pavlov in round $t+3$ is given by $A_{t+3}^{({\rm Y})} > P$, i.e., 
\begin{equation}
 (1-h)^2 < \frac{T-P}{P-S}. 
 \label{eq:h iq2}
\end{equation}
Equation~\eqref{eq:h iq1} implies Eq.~\eqref{eq:h iq2} because $(P-S)/(T-P)>0$ and $h>0$. 
Therefore, the two players become Pavlov in round $t+3$ if Eq.~\eqref{eq:h iq1} holds true. 

We assume that Eq.~\eqref{eq:h iq1} is violated. 
If Eq.~\eqref{eq:h iq2} is also violated, we obtain $A_{t^\prime}^{({\rm X})},\ A_{t^\prime}^{({\rm Y})} \leq P\ (t^\prime \geq t+3)$ such that the two players mutually defect until the occurrence of another action misimplementation. 
If Eq.~\eqref{eq:h iq2} is satisfied, the two players mutually defect in round $t+3$. 
Because $r_{t+2}^{({\rm Y})}=r_{t}^{({\rm X})}=T,\ r_{t+3}^{({\rm Y})}=r_{t+1}^{({\rm X})}=P$, and $A_{t+2}^{({\rm Y})}<P$, we obtain $P<A_{t+4}^{({\rm Y})}<A_{t+2}^{({\rm X})}$. 
Because $r_{t+2}^{({\rm X})}=r_{t}^{({\rm Y})}=S,\ r_{t+3}^{({\rm X})}=r_{t+1}^{({\rm Y})}=P$, and $A_{t+2}^{({\rm X})}>P$, we obtain $A_{t+2}^{({\rm Y})}<A_{t+4}^{({\rm X})}<P$. 
These two inequalities indicate that X and Y behave as st3 and st2 in round $t+5$, respectively. 
By repeating the same procedure with X and Y swapped, we obtain 
\begin{equation}
 A_{t+2}^{({\rm Y})}<A_{t+6}^{({\rm Y})}<P<A_{t+6}^{({\rm X})}<A_{t+2}^{({\rm X})}. 
\end{equation}
Therefore, we obtain 
\begin{equation}
 A_{t+2}^{({\rm Y})}<A_{t+2+4i}^{({\rm Y})}<P<A_{t+2+4i}^{({\rm X})}<A_{t+2}^{({\rm X})}\ \ \ (i \geq 1) 
 \label{eq:At iq}
\end{equation}
by induction. 
Equation~\eqref{eq:At iq} implies that the two players do not realize mutual cooperation if Eq.~\eqref{eq:h iq1} is violated. 

Therefore, an upper bound of $h$ for a pair of st2 players to turn into Pavlov is given by solving Eq.~\eqref{eq:h iq1} with equality. 

\clearpage
\clearpage
\clearpage
\clearpage
\clearpage
\clearpage
\clearpage
\clearpage
\clearpage
\begin{table}[htbp]
 \centering
  \caption{
Average payoff to a nonlearning player (row player) playing against an opponent nonlearning player (column player). 
We set $\beta=\infty,\ 0<\epsilon\ll 1,$ and $t_{\max}=\infty$.}
 \label{table:tab1}
 \begin{tabular}{|c||c|c|c|c|c|}
  \hline
  \makebox[4em][c]{} & 
  \makebox[4em][c]{st1} & 
  \makebox[4em][c]{st2} & 
  \makebox[4em][c]{st3} & 
  \makebox[4em][c]{st4} & 
  \makebox[4em][c]{st5} \\
  \hline\hline
    st1 & $\frac{R+T+S+P}{4}$ & $\frac{R+2S+3P}{6}$ & $\frac{R+T+S+P}{4}$ & $\frac{T+2S+P}{4}$ & $\frac{R+T+S+P}{4}$ \\
  \hline
    st2 & $\frac{R+2T+3P}{6}$ & $P$ & $\frac{R+2T+2P}{5}$ & $\frac{T+P}{2}$& $\frac{T+P}{2}$ \\
   \hline
    st3 & $\frac{R+T+S+P}{4}$ & $\frac{R+2S+2P}{5}$ & $R$ & $\frac{R+S+P}{3}$ & $\frac{R+T+S+P}{4}$ \\
   \hline
    st4 & $\frac{2T+S+P}{4}$ & $\frac{S+P}{2}$ & $\frac{R+T+P}{3}$ & $\frac{R+P}{2}$ & $\frac{R+P}{2}$ \\
   \hline
    st5 & $\frac{R+T+S+P}{4}$ & $\frac{S+P}{2}$ & $\frac{R+T+S+P}{4}$ & $\frac{R+P}{2}$ & $\frac{R+T+S+P}{4}$ \\
   \hline
 \end{tabular}
\end{table}
\end{document}